\documentclass[
preprint,
eqsecnum,
amsmath,amssymb, aps,
]{revtex4}

\newcommand{\be}{\begin{equation}}
\newcommand{\ee}{\end{equation}} 
\newcommand{\bea}{\begin{eqnarray}} 
\usepackage{graphicx}
\newcommand{\eea}{\end{eqnarray}}

\usepackage{epsfig}
\usepackage{bm} 
\usepackage{helvet}

\usepackage[dvipsnames]{xcolor}

\begin{document}

\title{Circulation Fluctuations of Elementary Turbulent Vortices }

\author{L. Moriconi$^{1}${\footnote{moriconi@if.ufrj.br}} and R.M. Pereira$^{2}${\footnote{rodrigomp@id.uff.br}}}
\affiliation{$^{1}$Instituto de F\'{i}sica, Universidade Federal do Rio de Janeiro, 21941-909, Rio de Janeiro, RJ, Brazil}
\affiliation{$^{2}$Instituto de F\'{i}sica, Universidade Federal Fluminense, 24210-346 Niter\'{o}i, RJ, Brazil}




\newtheorem{theorem}{\bf Theorem}[section]
\newtheorem{condition}{\bf Condition}[section]
\newtheorem{corollary}{\bf Corollary}[section]
\begin{abstract}
Thin vortex tubes, with core sizes within the dissipation range, profuse 
in a homogeneous and isotropic turbulent flow. Their intersections with an arbitrary plane define, as a mathematical construct, a dilute gas of localized, intermittently distributed, two-dimensional vortex spots. While their planar density fluctuations are described by a field-theoretical extension of log-normal single-point statistics, known as Gaussian multiplicative chaos (GMC), they carry circulations which are Gaussian-correlated throughout the inertial range. It is puzzling, then, to find that the circulations of individual vortices are fat-tailed distributed, an apparent paradox that we fix within the GMC framework. 
The solution, validated through the examination of direct numerical simulation data for a broad range of Reynolds numbers, unveils, as a surprising phenomenological result, an existing coupling between the circulation of vortex structures and the short-distance properties of their spatial distribution fluctuations at sub-Taylor microscales.
\end{abstract}





\maketitle

\section{Introduction}

Turbulence theory has been the stage of numerous scientific innovations over time, many resonating with or even preceding the introduction of important similar techniques in statistical physics. Notable examples are the Reynolds stress decomposition \cite{reynolds} vis-à-vis the mean field theory of magnetization \cite{curie,weiss}, the extension of statistical methods from particle systems to fields \cite{taylor,taylor2}, anomalous diffusion \cite{richardson,levy,mandelbrot}, the underlying motivation for the inception of functional integration by Wiener \cite{wiener, kac}, early discussions on scale invariance and anomalous scaling in physical models \cite{K41,O62,K62}, and the analogy between the scaling strategy of large eddy simulations (LES) \cite{smagorinsky} and the renormalization group approach to critical phenomena \cite{kadanoff}.
Other no less meaningful instances could also be added to the list.

An important lesson taken from this large variety of approaches is that significant advances in the understanding of turbulence have often been driven not by incremental refinements of some prescribed methodological framework, but by renewed cycles of inquiry or by the adoption of phenomenologically distinct viewpoints. Turbulence remains a mosaic of conceptions that have yet to be coherently integrated at a deeper level of modeling \cite{Jorg-Sreeni}. 

Within this broad context, the {\it{vortex gas model}} (VGM) of circulation statistics \cite{apol_etal,bounded_measures,mori_etal,mori_pereira,mori_etal2,iyer_mori}, the main focus of this work, provides a compelling bridge between two mainstream perspectives on homogeneous and isotropic turbulence (HIT). On one hand, it embraces the view that small-scale turbulent fluctuations arise from a multiplicative cascade process \cite{O62,K62}. On the other hand, it supports the structural approach, which posits that vorticity-containing structures are a relevant source of inertial range scaling phenomena. The structural line of thought is actually inspired by persuasive numerical simulations which strongly suggest that HIT could be depicted as a complex system of interacting vortex tubes \cite{orszag_etal, farge_etal, kaneda_etal, kaneda_etal2, burger_etal, afonso_etal}.

The fundamental physical observable upon which the VGM is based is the velocity circulation around an arbitrarily chosen oriented contour $C$,
\be 
\Gamma[C] \equiv \oint_C d x_i v_i ( x,t) \ . \  \label{gamma}
\ee
The essential motivation for modeling $\Gamma[C]$ comes from the Stokes theorem: circulation works as a mathematical probe of vortex structures and, therefore, is a most valuable tool for gathering insightful information about the statistical properties of HIT.

As a matter of fact, drawing on analogies with the loop calculus approach to quantum chromodynamics \cite{migdal1}, Migdal proposed in the mid-1990s that circulation could play a central role in the statistical theory of turbulence \cite{migdal2}. At that time, despite the partial success of embryonic attempts \cite{cao_etal}, numerical simulations -- the usual catalyser of progress in HIT -- were unable to comprehensively explore the statistics of turbulent circulation. Only recently, with the improvement of computational hardware, have larger turbulence datasets at much higher Reynolds numbers become available for post-processing, enabling a breakthrough in the study of circulation statistics. Major results that came to light are the approximate bifractality of circulation fluctuations \cite{Iyer_etal} and their dependence, after proper rescaling, on the minimal surface area bounded by the circulation contour $C$ \cite{Iyer_etal2,mori_PNAS}.

The VGM, so far restricted to the analysis of circulation around planar contours, is based on two general assumptions, namely
\vspace{0.2cm}

\noindent (i) that the turbulent circulation can be well approximated by the overall vorticity flux of thin elementary vortex tube structures that cross the planar domain enclosed by $C$;
\vspace{0.2cm}

\noindent (ii) that the probability of finding a vortex tube at some particular position of the flow depends on the local energy dissipation rate.
\vspace{0.2cm}

The cores of elementary vortices have linear dimensions around $10 \eta$ \cite{mori_etal,afonso_etal,ishihara_etal}, where $\eta$ is the Kolmogorov dissipation scale. Their circulations are furthermore assumed to be Gaussian correlated and polarized at short distances, while the energy dissipation field is independently described by Gaussian multiplicative chaos (GMC) theory \cite{GMC}, a formalism that provides a field theoretical generalization of the Obukhov-Kolmogorov (OK62) log-normal model of intermittency \cite{O62,K62}. The VGM has consistently reproduced or predicted several statistical properties of circulation, including the ones related to the small deviations of bifractality and fluctuations of the vortex density field \cite{bounded_measures,mori_pereira}.

One might nevertheless pinpoint an apparent modeling inconsistency of the VGM, having in mind the fact that the single-point statistics of the vorticity components is non-Gaussian \cite{kholmy_etal}: how can the circulations of elementary vortex structures be correlated through a position-dependent multivariate Gaussian distribution, if they should be individually represented as non-Gaussian random variables? This is the trick question that calls for a more careful analysis of the vortex gas modeling principles. 

This paper is organized as follows. Technical details of the VGM are briefly reviewed in Sec. II. In Sec. III, we put forward a solution, based on the GMC theory, for the modeling issues related to the fluctuations of elementary vortex circulations. The proposed solution is validated, in Sec. IV, through meticulous analyses of direct numerical simulation (DNS) data. Finally, in Sec. V, we discuss and summarize our findings, highlighting interesting directions for future research.

\section{Vortex Gas Model}

\begin{figure}[t]
\begin{center}
\hspace{0.0cm} \includegraphics[width=0.8\textwidth]{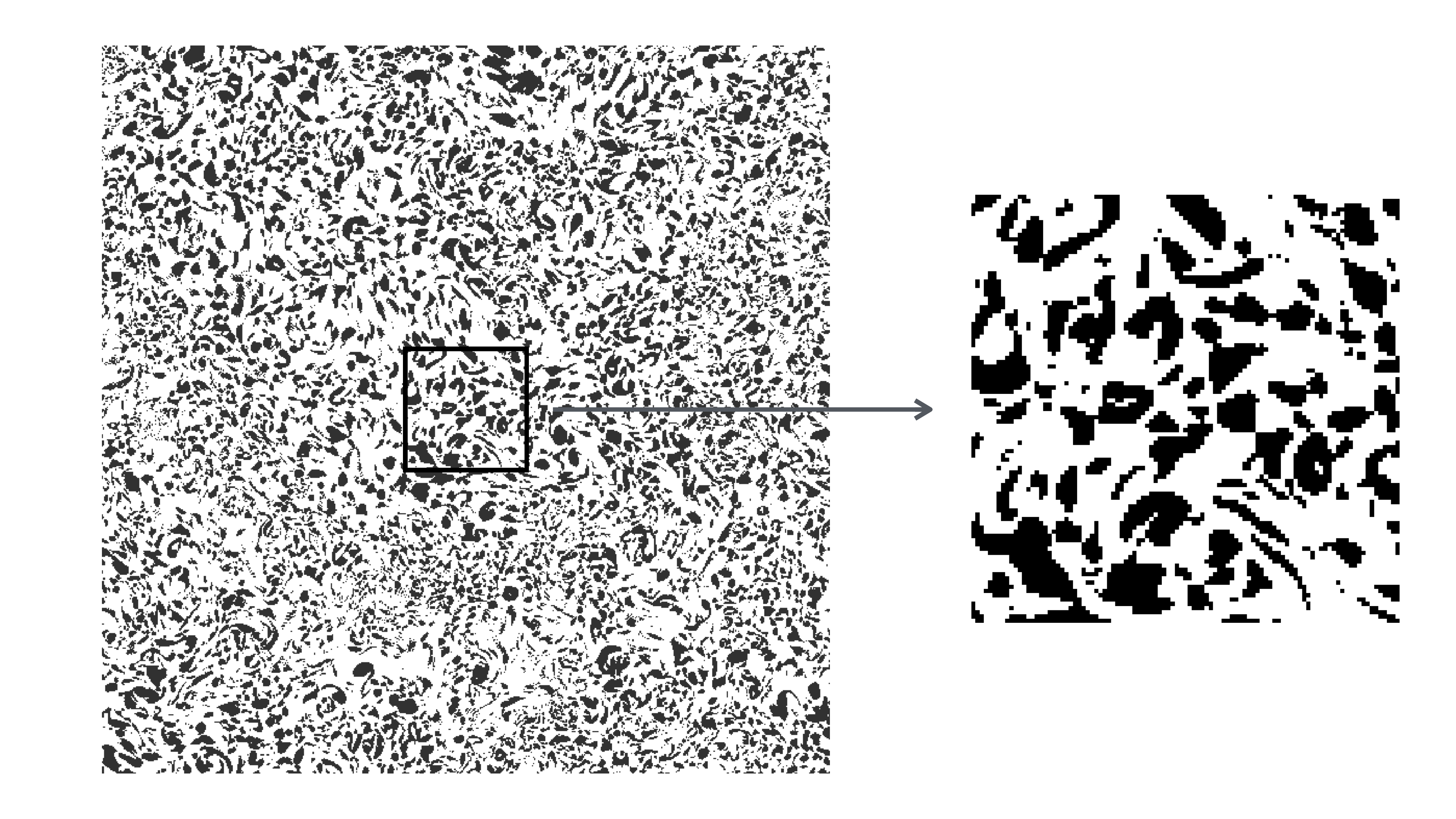}
\end{center}
\vspace{-0.0cm}
\caption{The granular pattern of dark spots in the left image represents cross-sections of vortex tubes intersected by a {\hbox{$600 \times 600$}} planar slice of a \(1024^3\) simulation domain (Taylor-scale Reynolds number \(R_\lambda = 433\); data obtained from the Johns Hopkins Turbulence Database~\cite{JHTD} and post-processed with the swirling strength criterion to identify vortex structures~\cite{zhou_etal}). The central \(100 \times 100\) region of the slice is magnified in the right-hand image.}
\label{spots}
\end{figure}

We summarize here the core technical content of the VGM. Let $x \in \mathbb{R}^2$ be a Cartesian parametrization of a planar domain embedded in a three-dimensional HIT flow. In the original version of VGM (to be critically revisited and improved in the next section) \cite{apol_etal, bounded_measures} we write
the circulation around a domain $\mathcal{D}$ bounded by a contour $C$ as
\begin{equation}
\Gamma[C] = \int_{\mathcal{D}} dN(x) \tilde \Gamma(x) \ , \ \label{gammadN}
\end{equation}
where
\be
dN(x) \equiv   d^2x \frac{\xi(x)}{\eta^2} \label{dN}
\ee
is the number of cross-sectioned vortex structures, each carrying an elementary circulation \(\tilde{\Gamma}(x)\), in a small area element \(d^2 x\) within the domain \(\mathcal{D}\).
In Eq.~(\ref{dN}), thus, $\xi(x)$ is the area number density of intersected vortex tubes, in units of the squared inverse Kolmogorov length scale. As a suggestive illustration, Fig.~\ref{spots} shows a typical planar slice of the flow, in the form of a dilute gas of small, compact vortex structures.

The continuum model given by Eq.~(\ref{gammadN}) has been applied for domain sizes $\mathcal{D}$ that range from the bottom of the inertial range up to the integral scales. At the smaller dissipative scales, it is necessary to refine it, in order to deal with the granular nature of vortex structures. This can be achieved while maintaining the continuum formulation, by introducing a point-process correction factor into the integrand of (\ref{gammadN}) \cite{apol_etal}.

The two-dimensional random fields $\xi(x)$ and $\tilde \Gamma(x)$ are taken to be completely independent. The vortex density turns out to scale as the square root of the energy dissipation field $\epsilon(x)$ \cite{mori_pereira}, a phenomenological point that can be quantitatively addressed within the language of GMC theory. More concretely, we have
\be
\xi(x) \equiv \xi_0 \exp [\sqrt{\pi \mu/2} \phi(x)] \propto \sqrt{\epsilon(x)/\epsilon_0} \ , \ \label{xi}
\ee
where $\epsilon_0 = \langle \epsilon(x) \rangle$, $\mu = 0.17 \pm 0.01$ is the intermittency exponent that rules the power law decay of the energy dissipation field correlator \cite{tang_etal}, $\phi(x)$ is a two-dimensional free scalar Gaussian field, characterized by the two-point correlation function
\be
\langle \phi(x) \phi(x') \rangle =  \frac{1}{(2 \pi)^2} \int
\frac{d^2k}{k^2} e^{ik \cdot(x-x')} \ , \ \label{phi-phi}
\ee
and $\xi_0$ is a dimensionless normalization constant.
The above integral is meant to be evaluated in the presence of natural ultraviolet and infrared cutoffs given, respectively, by $k_\eta = 1/\eta$ and $k_L = 1/L$, where $L$ is the integral scale of the flow.

The elementary circulation field $\tilde \Gamma(x)$ is, on its turn, assumed to be a Gaussian random field parametrized by a circulation scale $\tilde \Gamma_0$ and a scaling exponent $\alpha$. Its two-point correlation function is defined as
\be
 \langle \tilde \Gamma(x) \tilde \Gamma(x') \rangle = \tilde \Gamma_0^2 \frac{\eta^\alpha}{2 \pi \Gamma(\alpha)} \int d^2 k k^{\alpha-2} 
 e^{ik \cdot(x-x') - k\eta} \ , \ \label{gamma-gamma1}
\ee
so that  $\langle \tilde \Gamma^2 \rangle = \tilde \Gamma_0^2$, and
\be
 \langle \tilde \Gamma(x) \tilde \Gamma(x') \rangle \sim |x - x'|^{-\alpha} \ , \ \label{gamma-gamma2}
\ee
for $|x - x'| \gg \eta$. In (\ref{gamma-gamma1}), $\Gamma(\alpha)$ is the Euler-Gamma function of $\alpha$ (not to be confused with the circulation variable).

In order to set the value of $\alpha$, we first note, through a power counting analysis based on (\ref{gammadN}-\ref{gamma-gamma2}), that for a square contour of side $r$ within the inertial range scales, it holds
\be
\langle (\Gamma[C])^2 \rangle \sim r^{ 4 - \frac{\mu}{4} -\alpha} \ , \ \label{Gamma_sc1}
\ee
Alternatively, it is not too difficult to show,
from (\ref{gamma}), that
\be
\langle (\Gamma[C])^2 \rangle = 
\frac{1}{2} \oint_C d x_i \oint_C d x'_j \langle \delta v_i(x,x') \delta v_j(x,x') \rangle \ , \ \label{GammaGammav}
\ee
with
\be
\delta v_i(x,x') \equiv v_i(x) - v_i(x') \ . \ 
\ee
It is clear from (\ref{GammaGammav}), therefore, that the way how
$\langle (\Gamma[C])^2 \rangle$ scales is directly related to that
of the second order velocity structure functions \cite{frisch},
\be
D_{ij}(|x - x'|) \equiv \langle \delta v_i(x,x') \delta v_j(x,x') \rangle \sim |x- x'|^{\zeta_2} \ . \ \label{Dij}
\ee
Resorting again to power counting, we find, from (\ref{GammaGammav}) and
(\ref{Dij}), that 
\be
\langle (\Gamma[C])^2 \rangle \sim r^{ 2 + \zeta_2} \ . \ \label{Gamma_sc2}
\ee
Comparing $(\ref{Gamma_sc1})$ and $(\ref{Gamma_sc2})$, we get
\be
\alpha = 2 - \frac{\mu}{4} - \zeta_2 \ . \ 
\ee
Since $\zeta_2$ is only slightly off 2/3 and $\mu$ is small enough, it follows that {\hbox{$\alpha > 0$}} and, as a consequence, the integral in (\ref{gamma-gamma1}) converges in the infrared region.

The great usefulness of the GMC theory is that it yields closed expressions for the expectation values of source functionals as
\be
Z[j(x)] = \left \langle \exp \left [ \int d^2x j(x) \phi(x) \right ] \right \rangle  = \exp \left [ \frac{1}{2} \int d^2x d^2 x' j(x) j(x') 
\langle \phi(x) \phi(x') \rangle \right ] \ , \ \label{Z}
\ee
from which correlation functions related to the vortex density field $\xi(x)$ can be computed.

We point out, for the sake of completeness, that the scalar field $\phi(x)$ is only {\it{approximately}} free. The smaller are the fluctuations of $\phi(x)$, the better is the free field approximation. Larger fluctuations tend to be suppressed, and this affects the shape of the circulation probability distribution tails \cite{mori_pereira}. Also, the suppression effect leads to an upper bound limit for the values of the vortex density field $\xi(x)$, a situation which is phenomenologically connected with the observed short-distance repulsion between elementary vortices \cite{mori_etal}. In practical terms, the free-field assumption provides a good approximation for modeling circulation statistical moments $\langle (\Gamma[C])^q \rangle$ up to order $q \approx 6$ \cite{bounded_measures}.

\section{Elementary Vortices}\label{sec:vortices}

Notwithstanding the remarkable accuracy of the VGM in accounting for numerical observations, there is, as already alluded to in the introduction, a major puzzle concerning the general validity of one of its modeling ingredients: the single-point statistics of the elementary circulation fluctuations is non-Gaussian, even though its correlation functions factorize as Gaussian random fields for well-separated vortices ($|x_i-x_j| \gg \eta$) \cite{apol_etal,mori_etal2}.

It is peculiar, thus, that the VGM performs so well even for the smallest scales of the flow. Somehow, the vortex density and the elementary circulation fields should be coupled in a ``conspiracy" to keep (\ref{gammadN}) as a meaningful representation of circulation all the way through scales. 

To address the Gaussian (large scale) versus non-Gaussian (small scale) correlations of the elementary circulation field, we introduce an effective crossover length scale $\ell$ associated to the transition between the inertial and the dissipative ranges, as a way to model the distinct roles of scales in the definition of the scalar field $\phi(x)$. More precisely, we split $\phi(x)$ into low and high wavenumber modes as
\begin{equation}
\phi(x) = \phi_<(x) + \phi_>(x) \ , \ \label{phi_split}
\end{equation}
where the subscripts $<$ and $>$ refer, respectively, to the wavenumber domains
\begin{equation}
\frac{1}{L} \leq k \leq \frac{1}{\ell} 
\end{equation}
and
\begin{equation}
\frac{1}{\ell} \leq k \leq \frac{1}{\eta} \ . \ 
\end{equation}
It is clear that $\ell /\eta$ is in principle a Reynolds-number dependent quantity.
Neglecting non-linear self-interaction effects (which are important only for very extreme fluctuations), we may
write that
\begin{eqnarray}
&&\langle \phi_> (x) \phi_< (x') \rangle = 0 \ , \ \\
&&\langle \phi_> (x) \phi_> (x') \rangle = \frac{1}{(2 \pi)^2} \int_>
\frac{d^2k}{k^2} e^{ik \cdot(x-x')} \ , \ \label{phi>} \\
&&\langle \phi_< (x) \phi_< (x') \rangle = \frac{1}{(2 \pi)^2} \int_<
\frac{d^2k}{k^2}   e^{ik \cdot(x-x')} \ . \ \label{phi<}
\end{eqnarray}
Substituting (\ref{phi_split}) into (\ref{xi}), we obtain the decomposition of the (original) vortex density field, 
\begin{equation}
\xi(x) = \xi_0 \xi_<(x) \xi_>(x) \ , \ \label{xid}
\end{equation}
where
\bea
&&\xi_<(x) = \exp [\sqrt{\pi \mu/2} \phi_<(x)] \ , \ \\
&&\xi_>(x) = \exp [\sqrt{\pi \mu/2} \phi_>(x)] \ . \ 
\eea
Digging into the foundations of the VGM, we propose that the elementary circulations $\tilde \Gamma(x)$ of the original formulation of the VGM be replaced by 
\be
\bar \Gamma(x) \equiv \xi^\beta_>(x) \tilde \Gamma (x) \ , \ \label{xiscales}
\ee
where $\beta$ is an exponent yet to be determined. The circulation around a domain $\mathcal{D}$ can be written, thus, as
\be
\Gamma [C] = \sum_{x_i \in \mathcal{D}} \bar \Gamma(x_i) = \sum_{x_i \in \mathcal{D}} \xi^\beta_>(x_i) \tilde \Gamma(x_i) 
= \int_\mathcal{D} d^2 x  \sigma(x|\epsilon(x)) \xi^\beta_>(x) \tilde \Gamma(x) \ , \ \label{gammasigma}
\ee
where
\begin{equation}
\sigma(x|\epsilon(x)) = \sum_i \delta^2(x-x_i) 
\end{equation}
is the density of vortex structures conditioned upon the dissipation field $\epsilon(x)$. Assuming, now, when taking ensemble averages, that
\begin{equation}
\sigma(x|\epsilon(x)) \eta^2 = \xi_0 \xi_<(x) \xi^{1-\beta}_>(x) \ , \  \label{sigmaepsilon}
\end{equation}
we get, using (\ref{xid}) and (\ref{gammasigma}),
\be
\Gamma [C] = \frac{1}{\eta^2} \int_{\mathcal{D}} d^2x \xi_0 \xi_< (x) \xi_>(x) \tilde \Gamma (x) = \frac{1}{\eta^2}  \int_{\mathcal{D}} d^2x \xi(x) \tilde \Gamma (x) \ , \
\ee
so that we are back, after an alternative phenomenological line of reasoning, to the VGM formulation given by (\ref{gammadN}) and (\ref{dN}), as it should be.

To determine the exponent $\beta$ we resort to the numerical evidence that $\sigma(x|\epsilon(x)) \eta^2$ and $\xi(x)$ are the same in probability law \cite{mori_pereira}, 
\be
\sigma(x|\epsilon(x)) \eta^2 \overset{d}{=} \xi(x) \ . \ \label{sigma-xi}
\ee
Using (\ref{sigmaepsilon}) and the $\mathbb{Z}_2$ symmetry, 
$\phi_>(x) \rightarrow - \phi_>(x)$, of the related probability measure, we get
\be
\sigma(x|\epsilon(x)) \eta^2 = \xi_0 \xi_<(x) [\xi_>(x) ]^{1-\beta} 
\overset{d}{=} \xi_0 \xi_<(x) [\xi_>(x) ]^{\beta-1} \ , \ 
\ee
so that (\ref{sigma-xi}) holds for $\beta =0$ or $\beta =2$. We discard the $\beta = 0$ solution, once it would lead to the original version of the VGM. We take, from now on, {\hbox{$\beta = 2$}}. 


In the following, we proceed by examining the statistical properties (individual and collective) of the elementary vortices to assess whether they can effectively cope with the modeling challenges of the VGM and to formulate testable predictions suitable for numerical validation.
\\

\noindent {\bf{(i) Correlation Functions of $\bar \Gamma(x)$}}
\vspace{0.2cm}

Taking into account that $\xi(x)$ and $\tilde \Gamma(x)$
are independent, Eq.~(\ref{xiscales}) leads to the N-point correlation function
\be
    \langle \bar \Gamma(x_1) \bar \Gamma(x_2)\ \ ... \ \bar \Gamma(x_N) \rangle = G_{\tilde \Gamma}^{(N)}(x_1,x_2,...,x_N) G_{\xi}^{(N)}(x_1,x_2,..,x_N)   \ , \ 
\ee
where
\begin{equation}
    G_{\tilde \Gamma}^{(N)}(x_1,x_2,...,x_N) =   
\langle \tilde \Gamma(x_1) \tilde \Gamma(x_2)\ \ ... \ \tilde \Gamma(x_N) \rangle
\end{equation}
and
\begin{eqnarray}
    && \hspace{-0.8cm} G_{\xi}^{(N)}(x_1,x_2,...,x_N) =   
    \langle \xi^2_>(x_1) \xi^2_>(x_2) \ ...\ \xi^2_>(x_N) \rangle \nonumber \\
    && \hspace{-0.8cm} = \left \langle \exp \left [\gamma \sum_{i=1}^N \phi_>(x_i) \right ]  \right \rangle  = c^N \exp \left [ \gamma^2 \sum_{i=1}^{N-1}
    \sum_{j=i+1}^N \langle \phi_>(x_i) \phi_>(x_j) \rangle \right ] \ , \  \label{xiG}
\end{eqnarray}
with  $\gamma = \sqrt{2 \pi \mu}$ and
\begin{equation}\label{cdef}
c = \exp \left [ \frac{\gamma^2}{2} \langle \phi_>^2  \rangle \right ] 
= \left ( \frac{\ell}{\eta} \right)^\frac{\mu}{2} \ . \ 
\end{equation}
Notice that to derive (\ref{xiG}) we have made implicit use of (\ref{Z}), with the specific source field
\be
j(x) = \gamma \int_> \frac{d^2k}{(2 \pi)^2}
\exp(-i k x) \sum_{n=1}^N \exp(ikx_n) \ . \
\ee

Let us now assume that the vortices are spaced at distances significantly larger than the characteristic length scale $\ell$, that is, $|x_i - x_j| \gg \ell$ in (\ref{xiG}). Define the separation {\hbox{$|x_1 - x_2| \equiv r$}}.
We have
\begin{eqnarray}
&&\langle \phi_>(x_1) \phi_>(x_2) \rangle = \frac{1}{(2\pi)^2} \int_{1/\ell}^{1/\eta} 
\frac{dk}{k}  \int_0^{2 \pi} d \theta e^{ikr \cos(\theta)}  \nonumber \\
&&= \frac{1}{2\pi}
\int_{r/\ell}^{r/\eta} 
\frac{dk}{k} J_0(k) \approx \frac{1}{2\pi} \int_{r/\ell}^{r/\eta} 
\frac{dk}{k}  \frac{\cos(k - \pi/4)}{\sqrt{k}} \nonumber \\
&&\approx 
\frac{1}{2\pi} \left [ \frac{\cos(r/\ell + \pi/4)}{(r/\ell)^{3/2}} - \frac{\cos(r/\eta + \pi/4)}{(r/\eta)^{3/2}} \right ] \ . \
\end{eqnarray}
From this relatively fast decay of the $\phi_>$ correlators, we conclude that $G_{\xi}^{(N)}(x_1,x_2,...,x_N) \approx c^N$ and that, for points separated by large enough distances, correlations of $\bar \Gamma(x)$ behave effectively as the ones of a Gaussian random field (specifically, $c \tilde \Gamma(x)$), as predicted by the VGM and subsequently corroborated from a numerical study \cite{mori_etal2}.
\vspace{0.2cm}

\noindent {\bf{(ii) Statistical Moments of $\bar \Gamma(x)$}}
\vspace{0.2cm}

From (\ref{gamma-gamma1}) and (\ref{xiG}) we immediately obtain 
\begin{equation}
\langle | \bar \Gamma|^q \rangle 
= c^{q^2} \langle | \tilde \Gamma |^q \rangle = A_q c^{q^2} 
\tilde \Gamma_0^q \ , \ \label{moments}
\end{equation}
where, in terms of the Euler-Gamma function,
\be
A_q = \frac{2^\frac{q}{2}}{\sqrt{\pi}} \Gamma \left ( \frac{q+1}{2} \right )
\ee
is the usual combinatorial factor of Gaussian moments.
An interesting relation to explore in numerical validations is the quadratic dependence of
the log-moments on $q$, conveniently written as
 \begin{equation}\label{moments_parabola}
\ln( \langle | \bar \Gamma|^q \rangle /  A_q ) = q^2 \ln(c) + q \ln( \tilde \Gamma_0) 
 \end{equation}
or the generalized flatnesses of $\bar \Gamma$,
 \begin{equation}
\frac{\langle | \bar \Gamma|^q \rangle}{\langle | \bar \Gamma|^\frac{q}{2}  \rangle^2 } = 
\frac{\Gamma\left ( \frac{q+1}{2} \right ) }{\left [ \Gamma \left ( \frac{q+2}{4} \right ) \right ]^2} c^\frac{q^2}{2} \ . \ 
 \end{equation}
\vspace{0.2cm}

\noindent {\bf{(iii) Probability Distribution Functions of $\Gamma^\star$}}
\vspace{0.2cm}

Circulation probability distribution functions for arbitrary planar contours have been previously derived in the VGM \cite{apol_etal, mori_etal}. Here, instead, we obtain them for the standardized circulations $\Gamma^\star (x)$ of the elementary vortices, viz.,
\begin{equation}
\Gamma^\star (x) \equiv \frac{\bar \Gamma (x)}{\sqrt{\langle (\bar \Gamma)^2 \rangle}} = 
 X(x) Y(x) \ , \ \label{gammastar}
\end{equation}
where
\be 
X(x) = \frac{\xi_>^2(x) }{\sqrt{\langle \xi_>^4 \rangle}} \ , \
Y(x) = \frac{\tilde \Gamma(x)}{\sqrt{\langle \tilde \Gamma^2 \rangle}} \ . \ 
\ee
We have, therefore,
\begin{equation}
\langle (\Gamma^\star)^2 \rangle = 
\langle X^2 \rangle = \langle Y^2 \rangle =  1 \ . \ 
\end{equation}
While $Y$ is a Gaussian random variable, $X$ is a lognormal random variable that can be written as

\begin{equation}
X(x) = \exp \left ( Z(x) - \langle Z^2 \rangle \right ) \ , \  
\end{equation}
where $Z(x) = \gamma \phi_>(x) $.
We have
\begin{equation}
\sigma_z^2 \equiv \langle Z^2 \rangle  = \gamma^2 \langle \phi_>^2 \rangle 
= 2 \ln c = \mu \ln \left ( \frac{\ell}{\eta} \right )  \ . \ 
\end{equation}
As a consequence of (\ref{gammastar}), the probability distribution function of $\Gamma^\star$ turns out to be
\be 
\rho(\Gamma^\star) =  \langle \delta (XY - \Gamma^\star) \rangle_{X,Y} = \frac{1}{2 \pi \sigma_z} \int_0^\infty dY \frac{1}{Y^2}
 \exp \left [ -\frac{1}{2 \sigma_z^2} ( \ln Y + \sigma_z^2 )^2 -\frac{(\Gamma^\star)^2}{2 Y^2} \right ] \ . \  \label{pdf_gammastar}
\ee

\noindent {\bf{(iv) Vortex Density Fluctuations}}
\vspace{0.2cm}

The identification (\ref{sigma-xi}) of the vortex number density $\sigma(x|\epsilon(x))$ with the GMC field 
$\xi(x)$ gives, as straightforward consequence, that the density-density correlation function decays with distance 
as a power law with exponent $-\mu/4$,
\be
\label{sigma-corr}
\langle \sigma(x|\epsilon(x)) \sigma(x'|\epsilon(x')) \rangle = \langle \xi(x) \xi(x') \rangle \sim |x-x'|^{-\frac{\mu}{4}} \ , \ 
\ee
which provides an alternative way to measure the intermittency parameter $\mu$.

Of further interest is the coarse-grained vortex density, taken for a compact domain, as a square box of side $r$ within the inertial range,
\be
\xi_{CG}(r) \equiv \frac{1}{r^2}\int_\square
d^2x \sigma(x|\epsilon(x))
\overset{d}{=}  \frac{1}{\eta^2 r^2}\int_\square
d^2x \xi(r) \ , \ \label{xi-square}
\ee
which has statistical moments \cite{mori_pereira}
\be
\langle [\xi_{CG}(r)]^q \rangle \sim \left ( \frac{r}{L} \right )^{\frac{\mu}{8} q(1-q)} \ . \ \label{xi-square-moments}
\ee

In the same fashion as in the GMC theory for the Obukhov-Kolmogorov coarse-grained energy dissipation \cite{O62,K62}, $\xi_{CG}(r)$ is well approximated by a lognormal random variable. To avoid normalization issues, consider 
\be
\xi_{CG}^\star (r) \equiv \frac{ \xi_{CG} (r) }{ \langle  \xi_{CG} (r) \rangle}  \ , \  \label{xicg-star}
\ee 
which we expect to be approximately lognormal as well, with 
log-mean 
\be
\langle \ln \xi_{CG}^\star (r) \rangle =  -\frac{\mu}{8} \ln \left ( \frac{L}{r} \right )  \label{lmean}
\ee
and log-variance
\be
\langle [\ln \xi_{CG}^\star (r)]^2 \rangle - \langle \ln \xi_{CG}^\star (r) \rangle^2 = \frac{\mu}{4} \ln \left ( \frac{L}{r} \right ) \ . \ \label{lvar}
\ee

\noindent {\bf{(v) Physical Meaning of the Length Scale $\ell$}}
\vspace{0.2cm}

As it happens very often, solving a problem often brings new challenges to the table. The crucial solution step of splitting the energy dissipation field into fast and slow modes is carried out at a (in principle) Reynolds number dependent length scale $\ell$, which then calls for a phenomenological interpretation within the GMC description. Notice that the latter makes no direct reference whatsoever to vortex tubes or to the main dissipative structures (shear layers) of the flow. 

In this connection, the detailed study of Ishihara et al. \cite{ishihara_etal}, on the characterization of shear layers and vortex tubes in HIT, provides an interesting clue. It turns out, following that work, that the Taylor microscale $\lambda$ gives the typical thickness of the dissipation layers which are strongly correlated with the presence of vortex tubes. 

The applicability of the GMC theory to model vortex clusters as an assembly of the random Gaussian circulations $\tilde \Gamma(x)$ (as in the original formulation of the VGM) should thus be restricted to scales, likely to be lower-bounded by $\ell$, where
the circulation of elementary vortices decouple from their host vortex sheets.
We set, therefore, as a working hypothesis, that $\ell \approx \lambda$.
\vspace{0.2cm}

To summarize the essential results we have established so far: the improved version of the VGM is still described by a vortex density field $\xi(x) \overset{d}{=} \sigma(x|\epsilon(x)) \eta^2$ for elementary vortices which have random circulation $\bar \Gamma(x)$ defined by (\ref{xiscales}) with $\beta =2$. The single-point circulation statistics is non-Gaussian, but its correlation functions for well-separated points factorize like the ones of a Gaussian random field.

At this point, recalling that the factorization property derived in (i) has been in fact the specific subject of a previous validation study \cite{mori_etal2}, we drive our attention to the numerical scrutiny of the above items (ii-v).

\section{Numerical Validations}\label{sec:numerics}

\begin{figure}[b]
\begin{center}
\includegraphics[width=0.95\textwidth]{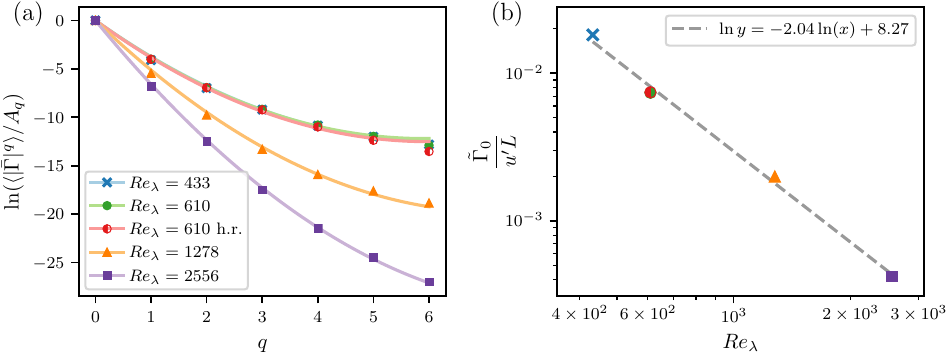}
\end{center}
\caption{(a) Statistical moments of individual vortex circulations \(\bar\Gamma\), normalized as in Eq.~(\ref{moments_parabola}), as a function of the moment order. Symbols: DNS data. Lines: parabolic predictions from Eq.~(\ref{moments_parabola}), with fitted linear coefficients. (b) \(\tilde\Gamma_0\) of Eq.~(\ref{moments_parabola}) obtained from the fitted coefficients, normalized by \(u'L\), as a function of \(Re_\lambda\) in logarithmic scale. Dashed: linear fit of the log-log data, revealing the -2 exponent predicted in Eq.~(\ref{gamma0scaling}).}
\label{fig:gamma_moms}
\end{figure}

Data analysis is carried out resorting to the DNS data made publicly available by the Johns Hopkins University Turbulence Database \cite{JHTD,JHTD2,JHTD3,JHTD4,JHTD5,JHTDBsite}. A total of five different datasets are used, with Taylor-based Reynolds numbers \(R_\lambda\) corresponding to 433, 610, 1278 and 2556. For \(R_\lambda=610\), two different simulations are analyzed: one in which the truncation wavenumber \(k_\text{max}\) obeys \(k_\text{max}\eta \approx 2.67\), and a higher resolution one, in which \(k_\text{max}\eta \approx 5.34\) (hereafter identified as ``h.r."). 

Following previous studies on the VGM \cite{mori_etal,mori_pereira,mori_etal2}, individual vortices are detected with the swirling strength criterion \cite{zhou_etal} applied to planar cuts extracted from the DNS. This should identify the intersections of the vortex structures and the plane. The main idea of the criterion, then, is to build a field \(\lambda(x)\) on the planar cut corresponding to any of the imaginary eigenvalues of the 2D velocity gradient tensor, and associate connected regions in which \(|\text{Im}(\lambda)|\) is non-zero to individual vortices. In an effort to suppress excessive noise, a threshold is usually employed, so, following \cite{mori_pereira}, we only identify regions where \(|\text{Im}(\lambda)|>\sigma_\lambda/8\), with  \(\sigma_\lambda^2\) being the variance of \(\ |\text{Im}(\lambda)|\). An example of a processed snapshot was already presented in Fig.~\ref{spots}.

For each simulation, an ensemble of planes is constructed by extracting multiple slices perpendicular to each Cartesian axis and, when available, at different time instants. The circulation of an individual vortex \(\bar\Gamma\) is determined by integrating the normal component of the vorticity over the area enclosed by the vortex, according to Stokes’ theorem applied to (\ref{gamma}). In this manner, an ensemble of realizations of \(\bar\Gamma\) is produced and then statistically analyzed. 
\begin{figure}[b]
    \begin{center}
    \includegraphics[width=0.6\textwidth]{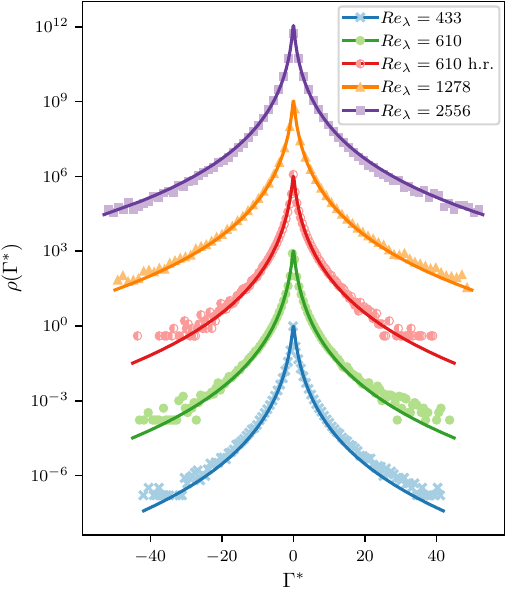}
    \end{center}
    \caption{Probability distribution function of the normalized circulation of individual vortices. Symbols: DNS data. Lines: analytical predictions of Eq.~(\ref{pdf_gammastar}). }
    \label{fig:gamma_pdf}
\end{figure}
We start by looking at the statistical moments of \(|\bar\Gamma|\) as a function of the moment order \(q\). In particular, we aim to verify Eq.~(\ref{moments_parabola}), which predicts a parabolic behaviour. We recall that, in our framework, the length scale \(\ell\) corresponds to the Taylor microscale \(\lambda\), so from (\ref{cdef}) one has \(c=(\lambda/\eta)^{\mu/2}\), given by the DNS parameters, and the only free parameter to be fitted in (\ref{moments_parabola}) is \(\tilde\Gamma_0\), or, equivalently, the first-degree coefficient of the parabola. 

In Fig.~\ref{fig:gamma_moms}(a), symbols show the left-hand side of (\ref{moments_parabola}) measured from the various DNS for \(q\) up to 6, while solid curves are parabolas whose second-degree coefficients are equal to \(\ln(c)\) and whose first-degree coefficients are obtained from a best-fit procedure.
\begin{figure}[b]       
    \begin{center}
    \includegraphics[width=0.6\textwidth]{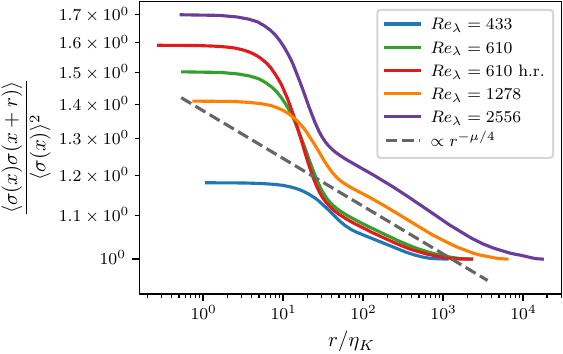}
    \end{center}
    \caption{Autocorrelation function of the numerically estimated vortex number density, compared to the scaling of Eq.~(\ref{sigma-corr}).}
    \label{fig:sigma_sigma}
\end{figure}
With these fitted linear coefficients, the quantity \(\tilde\Gamma_0\) from (\ref{moments}), and hence from (\ref{gamma-gamma1}), can be obtained for each simulation. We note that \(\tilde\Gamma_0\) represents a typical vortex circulation scale, and it has been argued 
that the most intense structures carry a circulation which scales as  \(u'\eta\) (with \(u'\) being the root mean square velocity fluctuation) \cite{afonso_etal}. Nevertheless, a computation of the ratio \(\tilde\Gamma_0/(u'\eta)\) for the different DNS we analyze here reveal no collapse, with values decreasing from 9 to 3 for increasing Reynolds numbers. This suggests that all structures, and not only the most intense ones, are important to recover the circulation statistics. With the typical vortex radius being of the order of \(4\eta\) \cite{mori_etal}, it is reasonable to expect the circulation scale to be of the order \(\tilde\Gamma_0\sim v_\eta\eta\), where \(v_\eta\) is a typical velocity fluctuation over a Kolmogorov length. In other words, since \(v_\eta\eta/\nu\sim 1\), \(\tilde\Gamma_0\) should scale with the viscosity \(\nu\), or \(\tilde\Gamma_0\sim u'L/Re \sim u'L/Re_\lambda^2\), which leads to
\be
\label{gamma0scaling}
\frac{\tilde\Gamma_0}{u'L}\sim Re_\lambda^{-2} \,.
\ee
This hypothesis is tested in Fig.~\ref{fig:gamma_moms}(b), which shows that the predicted power law behaviour \(Re_\lambda^{-2}\) is quite well observed from a linear fit of the log-log data.


Next, the probability distribution functions of the individual vortices circulations, normalized as in (\ref{gammastar}), are depicted in Fig.~\ref{fig:gamma_pdf}.
\begin{figure}[t]
    \begin{center}
    \includegraphics[width=0.9\textwidth]{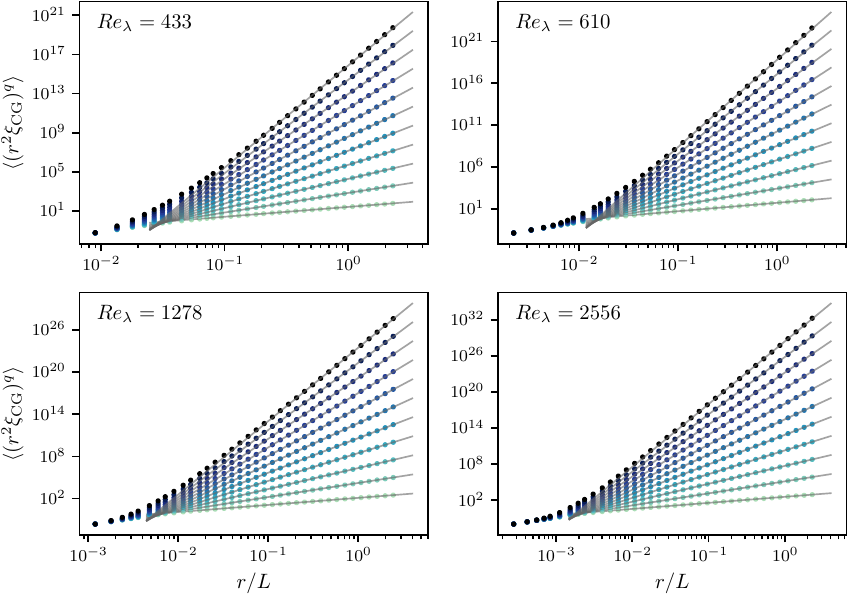}
    \end{center}
    \caption{Statistical moments of the coarse-grained vortex density as a function of the coarse-graining scale, for various moment orders \(q\). Lighter to darker colours represent \(q\) values ranging from 0.5 to 5.5, in steps of 0.5. Symbols: DNS data. Lines: scaling from Eq.~(\ref{xi-square-moments}) vertically shifted for a best fit to the data.}
    \label{fig:xiCG_moms}
\end{figure}
Once again, symbols represent the DNS data, following the same colour/marker type convention (which will be maintained from now on). Solid lines now show the analytical prediction of Eq.~(\ref{pdf_gammastar}). We emphasize that the excellent agreement observed here is obtained without any fitting procedure whatsoever.

We now turn to the study of the vortex number density distribution, identified, as discussed in Subsec.~\ref{sec:vortices}(iv), with the GMC field \(\xi(x)\). In our numerical study, it must be estimated from the post-processed DNS data. For this purpose, we have explored two different approaches. First, we computed the Voronoi cells \cite{okabe} around each vortex center, which consist of the set of points on the plane which are closer to a given vortex than it is to all the others. The local density \(\sigma(x)\) inside each cell is then estimated as the inverse area of the cell. A much more efficient approach, though, is a kernel density estimation, in which one sums Gaussian functions with a given width and centered at each vortex positions, to build a density function. This is efficiently done by convolving a Gaussian with an indicator function on the plane (which is 1 where a vortex center exists and 0 elsewhere). Vortex positions must be rounded to the computational grid, but this had no observable effect. The drawback of this approach is that it requires an additional parameter, known as the bandwidth (the width of the Gaussian function), however, we simply employed the typical vortex diameter 8\(\eta\), which proved to be a fine choice (we briefly comment on this below). Both approaches gave very similar results for the lower Reynolds numbers, where the Voronoi approach is more feasible, hence, for this reason, we kept the kernel density estimation for the analysis shown here.
\begin{figure}[b]
    \begin{center}
    \includegraphics[width=0.96 \textwidth]{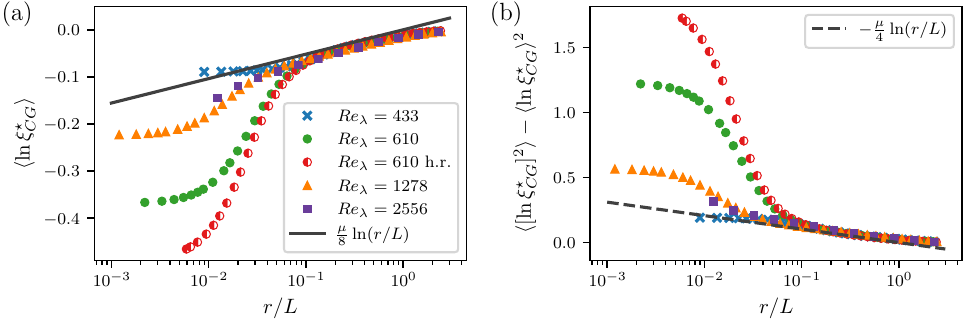}
    \end{center}
    \caption{(a) Log-mean and (b) log-variance of the coarse grained vortex density (normalized according to Eq.~(\ref{xicg-star})) as a function of the coarse-graining scale. Symbols: DNS data. Solid line: prediction from Eq.~(\ref{lmean}). Dashed line: prediction from Eq.~(\ref{lvar}).}
    \label{fig:logxiCG_moms}
\end{figure}
%

Fig.~\ref{fig:sigma_sigma} first verifies the relation (\ref{sigma-corr}), depicting the density autocorrelation function. 
No scaling range is seen for small scales (\(r\lesssim 40\eta\)), due to small scale regularization and the effect of the bandwidth parameter, but a reasonable scaling range emerges over intermediate scales, closely following Eq.~(\ref{sigma-corr}). These plots were also used to check the bandwidth choice: we verified that smaller bandwidths led to poorer results or even to a complete loss of the scaling range. Larger bandwidths, on the other hand, would shorten the scaling range, with no quality gain. Remarkably, a bandwidth right around the typical vortex diameter did indeed exhibit the largest scaling ranges in all cases.
 
Fig.~\ref{fig:xiCG_moms} displays the statistical moments of the coarse-grained vortex density as a function of the coarse-graining scale, along with the scalings predicted by Eq.~(\ref{xi-square-moments}). Different colours represent different moment orders \(q\), from \(q=0.5\) up to 5.5. Solid lines are obtained by taking the scaling from Eq.~(\ref{xi-square-moments}) and by fitting the proportionality constant in the region \(0.2\leq r/L\leq 1.5\) (this, of course, corresponds to fitting the vertical shift in the log-log plot).
    
Finally, Eqs.~(\ref{lmean}) and (\ref{lvar}) are verified in Fig.~\ref{fig:logxiCG_moms}. It shows the mean and the variance of the log of the coarse-graining vortex density, normalized as in Eq.~(\ref{xicg-star}), as a function of the coarse-graining scale. 

It is worth noting that, as with the circulation probability distributions in Fig.~\ref{fig:gamma_pdf}, the scaling results of Figs.~\ref{fig:sigma_sigma} –\ref{fig:logxiCG_moms} closely corroborate the VGM’s predictions without any fitting procedure.

\section{Discussion}

We have improved the phenomenological basis of the VGM of turbulent circulation by redefining the statistical properties of the circulation carried by elementary vortices. A relevant point to keep in mind is that while the reach of the VGM has been now broadened, the modeling features of its original version are all preserved across the inertial range down to the Kolmogorov dissipation scales.

The ``Gaussian puzzle" related to the original formulation of the VGM is implied by the (previously validated \cite{mori_etal2}) assumption that the multi-point correlation functions of the elementary circulation field factorize for different points, as is the case for a Gaussian random field. This seems to contradict the fact that vorticity Cartesian components, in the same fashion as general velocity gradients, are well known not to be Gaussian random variables \cite{kholmy_etal}. 

We note, in principle, that due to the intermittent spatial distribution of vortex structures, there could be a coexistence of Gaussian circulation statistics for the elementary vortices alongside non-Gaussian statistics for the vorticity components. This hypothesis, however, is promptly ruled out by observational evidence: the probability distribution functions of the elementary circulations are indeed fat-tailed.

Rephrasing the technical developments introduced in this work in a somewhat loose way, we have observed that when the spatial fluctuations of the energy dissipation field are decomposed into fast and slow modes, the fast modes can be absorbed into the definition of the elementary circulation fluctuations. As a result, the redefined vortex density field retains the same statistical properties as those postulated in the original formulation of the VGM. In this proposed resolution of the Gaussian puzzle, the GMC theory of the multifractal energy dissipation field has proven to be absolutely instrumental.

The reformulated VGM leads to a number of predictions which were carefully corroborated from the analysis of DNS data in Sec.~\ref{sec:numerics}. In particular, we call attention to the emblematic verification of the closed analytic expression for the probability distribution function of the normalized elementary circulation $\Gamma^\star$, given in (\ref{pdf_gammastar}), without resorting to any fitting procedure, as shown in Fig.~\ref{fig:gamma_pdf}.

As a drive for further research, it would be interesting to investigate the structure of the VGM in the context of LES. One could expect small-scale clusters of polarized vortex tubes to play the role of elementary vortices in the filtered Navier-Stokes dynamics. That suggests that there should be a trade-off, as the dynamics is progressively coarse-grained, between the vortex density field (taken as random measure) and elementary circulation fluctuations, a renormalization phenomenon that could impact the design of alternative LES closures.

\acknowledgments{The Gaussian puzzle in the original formulation of the VGM -- the triggering motivation of this paper -- was actually emphasized (and urged for a solution) by Luca Biferale on the occasion of the 2023 ICTS-Bangalore meeting entitled ``Field Theory and Turbulence". We thank the Interdisciplinary Center for Fluid Dynamics at UFRJ for computational resources.
This work was partially supported by the Brazilian National Council for Scientific and Technological Development (CNPq).}


\end{document}